# Development of a subwavelength grating vortex coronagraph of topological charge 4 (SGVC4)

Christian Delacroix*[a], Olivier Absil[a], Brunella Carlomagno[a], Pierre Piron[a], Pontus Forsberg[b], Mikael Karlsson[b], Dimitri Mawet[c], Serge Habraken[a] and Jean Surdej[a]

[a]Department of Astrophysics, Geophysics and Oceanography, University of Liège, 17 allée du Six Août, B-4000 Sart Tilman, Belgium;
[b]Ångström Laboratory, Uppsala University, Lägerhyddsvägen 1, SE-751 21 Uppsala, Sweden;
[c]European Southern Observatory, Alonso de Córdova 3107, Vitacura 7630355, Santiago, Chile

**ABSTRACT**

One possible solution to achieve high contrast direct imaging at a small inner working angle (IWA) is to use a vector vortex coronagraph (VVC), which provides a continuous helical phase ramp in the focal plane of the telescope with a phase singularity in its center. Such an optical vortex is characterized by its topological charge, i.e., the number of times the phase accumulates $2\pi$ radians along a closed path surrounding the singularity. Over the past few years, we have been developing a charge-2 VVC induced by rotationally symmetric subwavelength gratings (SGVC2), also known as the Annular Groove Phase Mask (AGPM). Since 2013, several SGVC2s (or AGPMs) were manufactured using synthetic diamond substrate, then validated on dedicated optical benches, and installed on 10-m class telescopes. Increasing the topological charge seems however mandatory for cancelling the light of bright stars which will be partially resolved by future Extremely Large Telescopes in the near-infrared. In this paper, we first detail our motivations for developing an SGVC4 (charge 4) dedicated to the near-infrared domain. The challenge lies in the design of the pattern which is unrealistic in the theoretically perfect case, due to state-of-the-art manufacturing limitations. Hence, we propose a new realistic design of SGVC4 with minimized discontinuities and optimized phase ramp, showing conclusive improvements over previous works in this field. A preliminary validation of our concept is given based on RCWA simulations, while full 3D finite-difference time-domain simulations (and eventually laboratory tests) will be required for a final validation.

## 1. INTRODUCTION

During the past twenty years, detections of extrasolar planets have flourished and grown exponentially, reaching almost 1800 confirmed exoplanets as of today. This number doubles every two or three years. The most recent discoveries made by the Kepler mission have revealed the first Earth-sized exoplanet situated in the habitable zone.[1] Imaging techniques have also come a long way. Since the first direct detection of a planetary-mass object orbiting a brown dwarf in 2004, some stunning exoplanet pictures have been captured under special circumstances of moderate contrast and/or angular separation,[2–8] thanks to the advent and continuous improvement of adaptive optics systems and data reduction methods. In this broad context, a high contrast coronagraph with a very small inner working angle (IWA) would have a key role to play in order to provide the means necessary for imaging Earth-sized planets, and to try answering the recurring question of the possible presence of life outside the solar system.

The high contrast and small IWA combo can be achieved with a phase mask coronagraph, such as the Four Quadrant Phase Mask (FQPM),[9] optimized so that destructive interferences send all the diffracted starlight outside the telescope aperture where it is intercepted by a diaphragm called Lyot stop. The main drawback of the FQPM is the transitions between the quadrants, causing phase discontinuities. In addition, the perfect alignment of the four quadrants is hard to manufacture. In order to avoid the tedious task of aligning the four quadrants of the FQPM, and to suppress the transition zones which cause a loss of information, a new solution was proposed by our team a few years ago. It is a broadband coronagraph called the Annular Groove Phase Mask (AGPM).[10] The idea is to replace the phase-shift steps by a continuous phase-ramp, thereby synthesizing a vectorial vortex. Vector Vortex Coronagraphs (VVCs) are known for their outstanding coronagraphic properties: small IWA (down to 0.9 $\lambda/D$), high throughput, clear off-axis 360° discovery space, and simplicity of fabrication.

*delacroix@astro.ulg.ac.be; phone +32 43669735; ago.ulg.ac.be

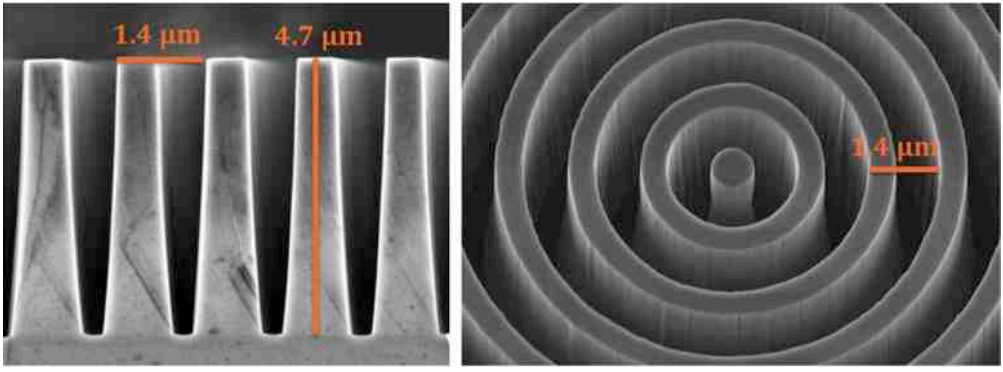

**Figure 1.** SEM-micrographs of a diamond AGPM dedicated to the L band. **Left**: Cross sectional view of the grooves on a cracked sample. **Right**: center of the AGPM.

To produce a broadband phase shift, we use the dispersion of form birefringence of subwavelength gratings (SGs). These are micro-optical structures with a period $\Lambda$ smaller than $\lambda/n$, $\lambda$ being the observing wavelength and $n$ the refractive index of the grating substrate. One can employ these SGs to fabricate circularly symmetric half-wave plates into a diamond substrate,[11] and hence produce an AGPM (see Fig. 1). The manufacturing method and working principle are largely explained in previous publications.[12–14] Recently, several broadband mid-infrared AGPMs were successfully manufactured based on SGs and validated on a coronagraphic test bench at the Observatoire de Paris, where a peak rejection up to 500:1 was measured over the entire L band.[15] Moreover, our components were installed on three infrared cameras: VLT-VISIR, VLT-NACO and LBT-LMIRCam.[16–19] The main advantage of the SG technology over other techniques such as Liquid Crystal Polymers (LCP)[20,21] or Photonic Crystals (PC)[22] is the bandwidth. LCP- and PC-VVCs can be affected by absorption at wavelengths beyond the near-infrared (H band centered at ~1.65μm, K band ~2.2μm). Furthermore, single-layer LCP VVCs are not intrinsically achromatic.

In this paper, we present our development of the next generation subwavelength grating vortex coronagraph (SGVC). We first detail our motivations for increasing the so-called topological charge in order to obtain a broader extinction, reducing the sensitivity to low-order aberrations (Sect. 2). We then compare different possible design solutions to reduce the issue of phase discontinuities and minimize their impact (Sect. 3), and we present their performance validation using RCWA simulations (Sect. 4). Finally, we conclude with the perspectives for 3D rigorous numerical simulations using a finite-difference time-domain (FDTD) method.

## 2. MOTIVATIONS FOR A CHARGE-4 VECTOR VORTEX CORONAGRAPH

Vector vortex coronagraphs provide a continuous helical phase ramp, varying azimuthally around the optical axis. This phase ramp can be written as $e^{il\theta}$, where $\theta$ is the focal plane azimuthal coordinate, and $l$ the (even) vortex topological charge, i.e., the number of times the geometric phase $\phi_p$ (Pancharatnam phase)[23] accumulates $2\pi$ along a closed path $s$ surrounding the phase singularity:

$$l = \left(\frac{1}{2\pi}\right) \oint \nabla \phi_p ds . \qquad (1)$$

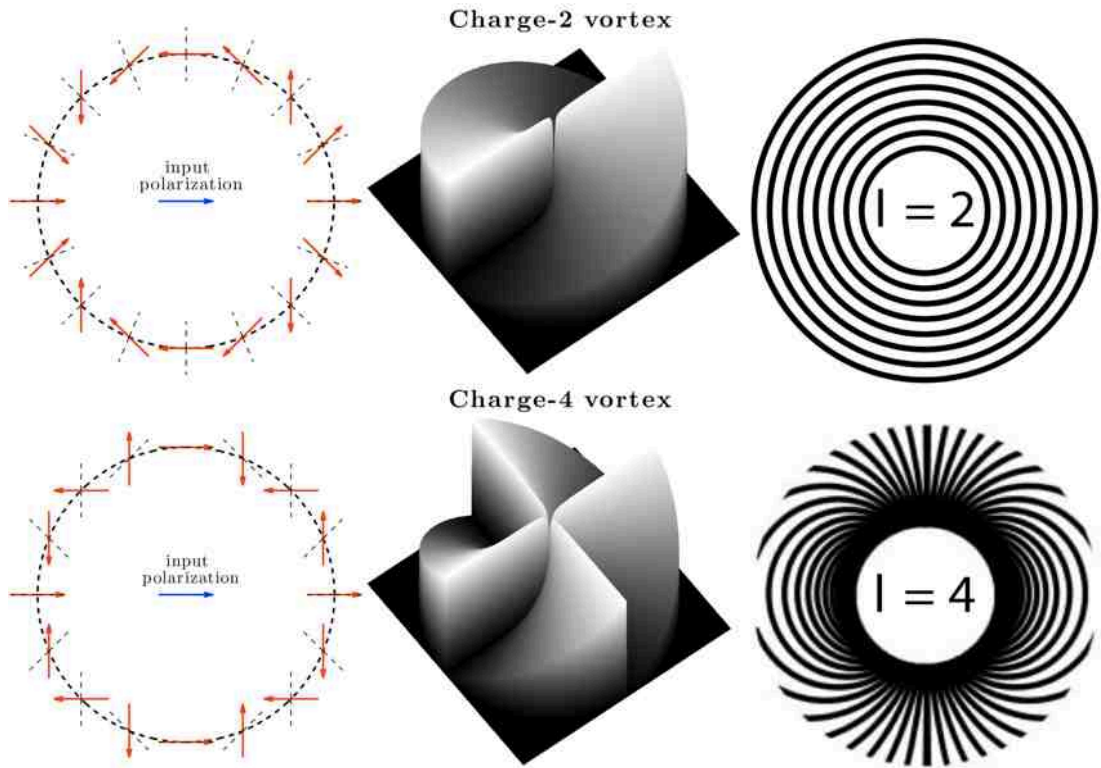

**Figure 2.** Working principle of a vector vortex coronagraph (VVC) of topological charge $l=2$ (**top**) and $l=4$ (**bottom**). **Left**: Illustration of a rotating half-wave plate (HWP). The optical axis orientation $\omega$ is represented with dashed lines, and it rotates about the center as $\omega=\theta l/2$, where $\theta$ is the azimuthal coordinate. The HWP effect is to rotate the input polarization (central arrow) by $-2\alpha$ where $\alpha$ is the angle between the incoming polarization direction and the optical axis orientation $\omega$. **Middle**: The VVC azimuthal phase ramp obtained equals $2\pi l$. **Right**: Binary grating geometry with constant line width. Only a charge-2 vector vortex coronagraph possesses the required circular symmetry for use with subwavelength gratings (SGVC2 aka AGPM), which permit achromatization. For any other charge $l \neq 2$, the grating period is locally larger than the subwavelength limit.

The present work focuses on the thorough study of second-generation vortex coronagraphs, by increasing the so-called topological charge $l$ of the vortex. Up to now, we have always manufactured subwavelength grating vortex coronagraphs of charge 2 (SGVC2 aka AGPM) for two reasons. Firstly, the SGVC2 allows smaller IWA ~0.9 $\lambda/D$ compared to an SGVC4 ~1.6 $\lambda/D$, approximately (depending on the optics setup). Secondly, its technical simplicity makes it easier to fabricate, and it is the only case to allow the required symmetry for the grating parameters to remain constant over the whole surface (see Fig. 2), thereby ensuring both the phase shift stability and the subwavelength regime. However, higher charges are mandatory to reduce the sensitivity to low order aberrations (tip-tilt, focus, coma, astigmatism). And more importantly, with the increased resolution of future ELTs, the performance of near-infrared coronagraphs will be affected by the finite size of nearby stars, which can be mitigated by the use of higher topological charge vortices.

To illustrate the benefits of increasing the topological charge, we have performed simulations of the sensitivity of the future E-ELT, as shown in Fig 3. These sensitivity limits only take into account the residual starlight leaking through the coronagraph due to the finite size of the star, and the contribution of the thermal background (adaptive optics residuals are not included here). These plots illustrate that an L-band instrument will be background-limited outward 1.6 $\lambda/D$ for almost all nearby stars ($L>1$), while a K-band instrument would be limited by the imperfect cancellation of stellar light for stars brighter than $K=7$. The use of a charge-4 vortex coronagraph can mitigate this limitation, albeit at the price of a degraded inner working angle (1.6 $\lambda/D$ instead of 0.9 $\lambda/D$ for the charge-2 vortex).

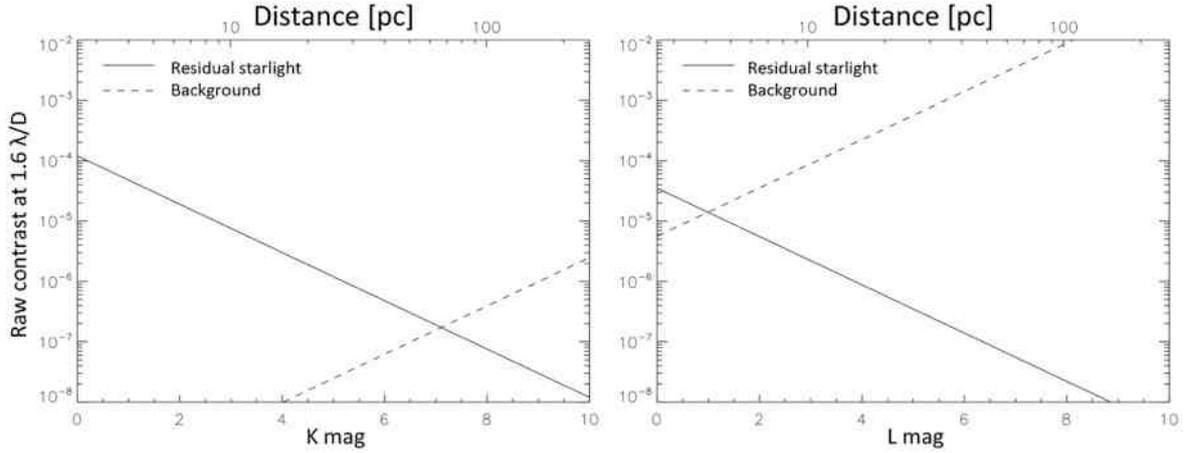

**Figure 3.** Sensitivity limit of an E-ELT instrument at 1.6 $\lambda/D$ from the star, using a charge-2 vector vortex coronagraph (such as an SGVC2, aka AGPM). The residual starlight comes from the finite size of the star. **Left**: K-band instrument. **Right**: L-band instrument.

## 3. NUMERICAL DESIGN OF AN SGVC4

### 3.1 Design of an SGVC4 using straight lines

In order to remain in the subwavelength regime, one must keep a grating period $\Lambda$ smaller than $\lambda/n$, $\lambda$ being the observing wavelength and $n$ the refractive index of the grating substrate. As shown in Fig. 2, the main limitation with a charge-4 continuous design is that, to reach the desired continuous optical axis orientation $\omega=2\theta$, the grating period must be locally larger than the subwavelength limit. To address this issue, one solution was proposed by Niv et al.[24] It consists in a discrete design where each discretization region (or slice) is a portion of $2\pi/N$, where $N$ is the number of slices. Each region is characterized by the same constant grating period $\Lambda$ (see Fig. 4).

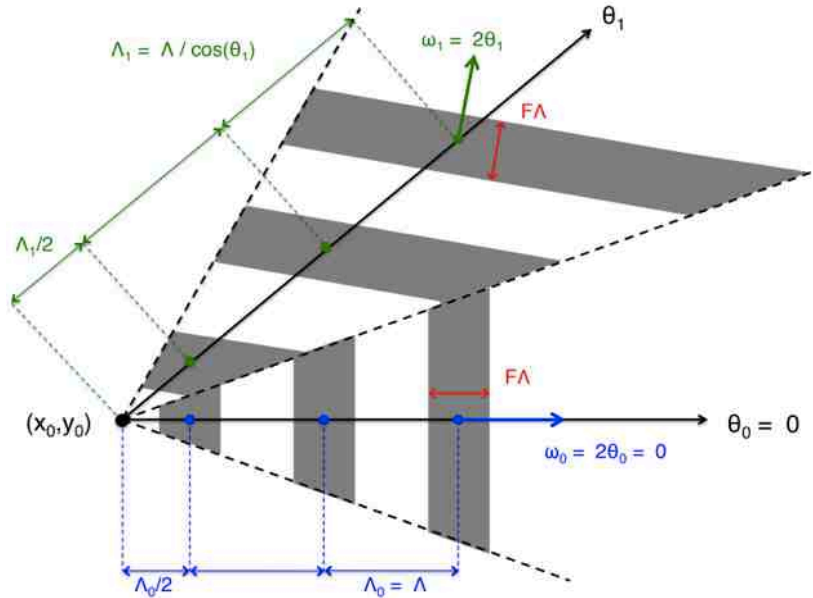

**Figure 4.** Geometrical illustration of the design of an SGVC4 with constant filling factor $F$, using straight lines. The whole component is discretized into $N$ portions of $2\pi/N$. The period $\Lambda$ and the line width $F\Lambda$ are identical in each region. To reach the desired optical axis orientation $\omega_i$, the radial period must vary as $\Lambda_i=\Lambda/\cos(\theta_i)$, $\theta_i$ being the azimuthal central coordinate of the i[th] region.

While computing the design shown in Fig. 4, we noticed that it presents several limitations.

(i) The discretization step being constant ($2\pi/N$) over the whole pattern, the number of portions seems unnecessarily large along the x-axis where $\omega_i$ is close to $\theta_i$, and hence $\Lambda_i$ is close to $\Lambda$. In contrast, we might want a fine discretization along the y-axis where $\Lambda_i$ becomes very high.

(ii) The local orientation of the optical axis is constant over each discretization region, preventing the polarization from continuously turning. This means that the desired phase ramp is also not perfectly continuous.

(iii) To get a phase ramp as close as possible to the desired continuous one, we need a large number of slices (typically $N=64$)[25] at the cost of an increased number of grating discontinuities.

Our analysis led us to explore a new design of an SGVC4, that could completely overcome the issues (i) and (ii), and highly alleviate the issue (iii).

### 3.2 Optimized design of an SGVC4 using curved lines

As mentioned before, it is geometrically not possible to obtain a continuous binary geometry for an SGVC4, with a constant grating line width (as shown previously in Fig. 2), and without forcing the period to be locally larger than the SG limit. In fact, a continuous pattern requires a varying period $\Lambda(\theta)$. Hence, as long as we impose a constant line width $F\Lambda$, we will need a varying filling factor $F$, which completely degrades the HWP performance, and therefore that of the vortex. As long as the manufacturing is forcing us to a constant etching depth over the whole pattern, it is forcing us to a constant filling factor.

With these conditions in mind, we came up with a new discrete design, using curved lines, variable in $F\Lambda$. We impose a constant $F$ and a maximal SG limit for $\Lambda$, at the price of a decreasing $\Lambda$ (and hence a decreasing $F\Lambda$) towards the y-axis, as illustrated in Fig. 5. With this optimized design, the discretization step is not constant over the whole pattern. In fact, when we compare both designs (lines and curves, see Fig. 6), we see that the number of discretization regions close to the x-axis is highly reduced with the new design, and the grating pattern close to the y-axis is much more precisely defined. But most of all, the orientation of the optical axis $\omega$ is rigorously defined everywhere over the pattern, which is necessary to obtain a continuous phase ramp of the vortex.

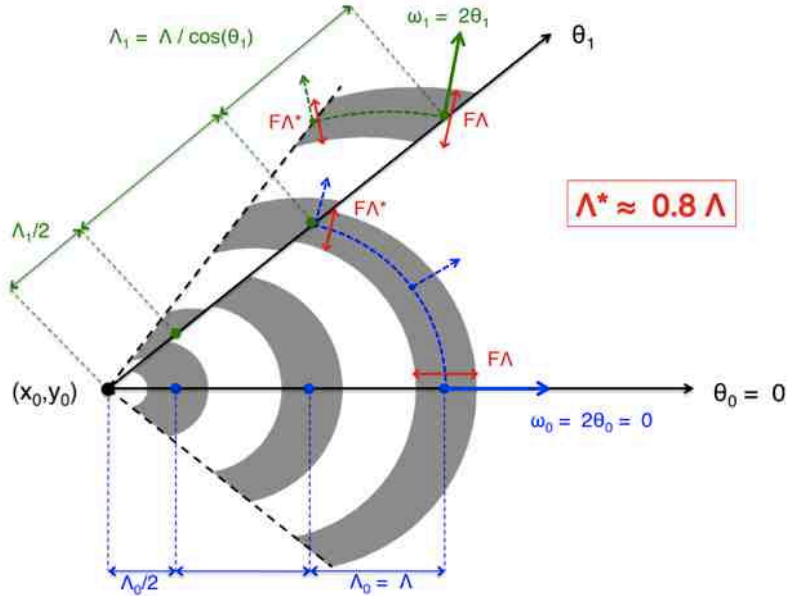

**Figure 5.** Geometrical illustration of the design of an SGVC4 with constant filling factor $F$, using curved lines. The optical axis orientation $\omega_i$ is rigorously defined everywhere, which implies that the period and the line width are decreasing towards the y-axis. The minimal period is fixed at $\Lambda^*=0.8\times\Lambda$. This condition defines the azimuthal coordinates $\theta_i$ for each region of discretization.

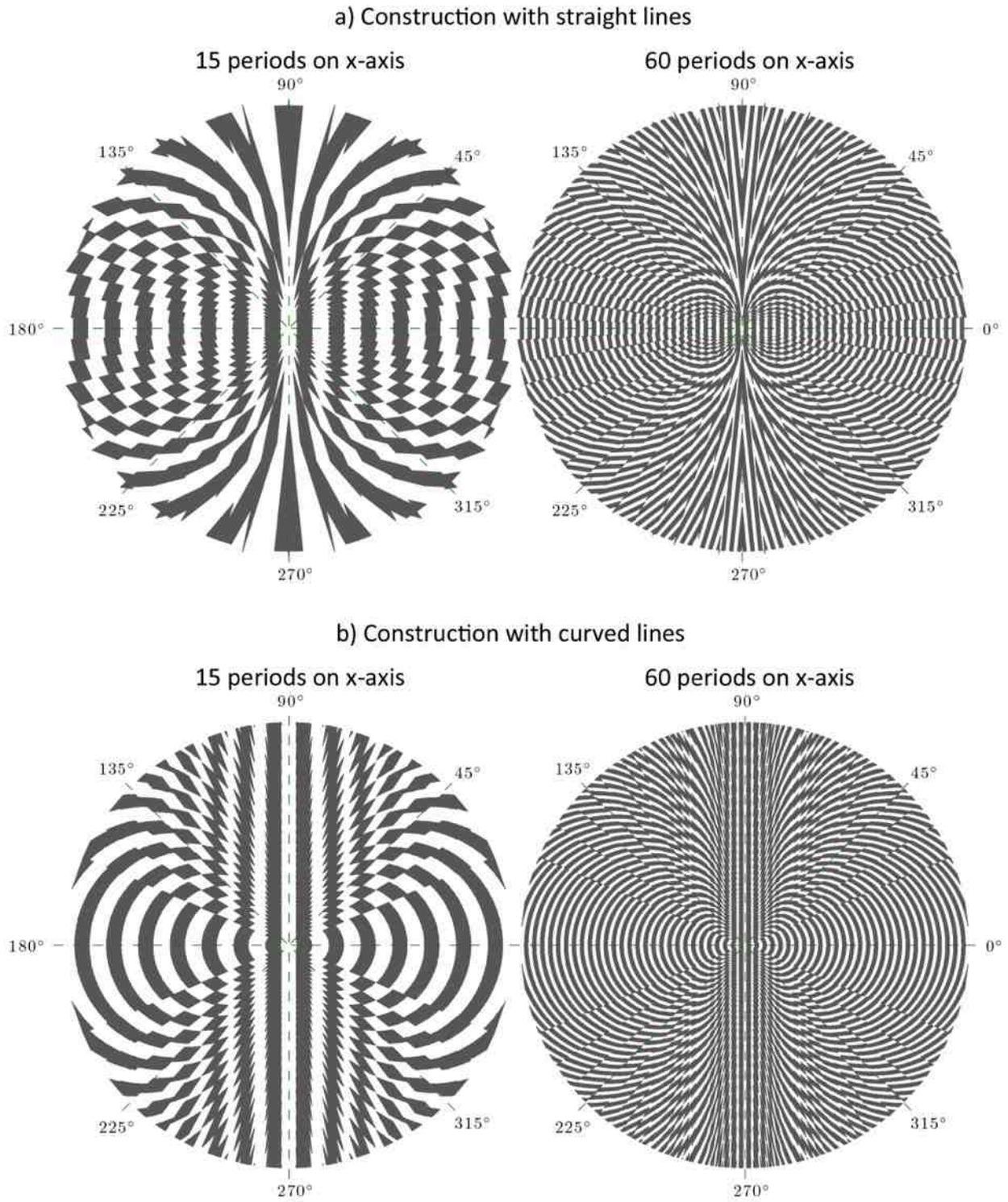

**Figure 6.** Comparison of two possible designs of an SGVC4, using either **(a)** straight lines or **(b)** curved lines. **Left**: close-up view of the central region of the coronagraph, with 15 periods on the x-axis. **Right**: wider overview, with 60 periods on the x-axis.

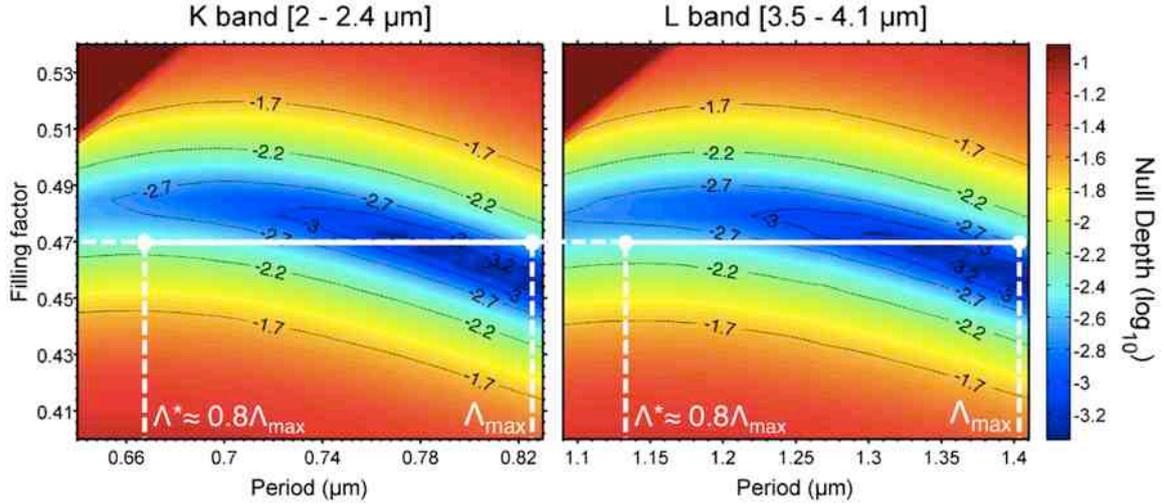

**Figure 7.** RCWA multi-parametric simulations. Theoretical mean null depth (logarithmic scale) over the whole K band (**left**) and L band (**right**), as a function of the period $\Lambda$ and the filling factor $F$. The grating thickness is kept constant. The optimal mean null depth (blue region) is situated at $F\sim0.47$ and $\Lambda$ ranging from $0.8\times\Lambda_{max}$ to $\Lambda_{max}$.

## 4. PERFORMANCE VALIDATION USING RCWA SIMULATIONS

We can evaluate the extinction performance of an SGVC by performing realistic numerical simulations using the rigorous coupled wave analysis (RCWA), which resolves the Maxwell equations in the frequency domain and gives the entire diffractive characteristics of the studied structure.[26] These analyses are however only rigorous in the 2D domain, and assuming that the SG is parallel and infinite, which is not the case for an SGVC. At this stage, we shall assume that these hypotheses are satisfied, knowing that our previous RCWA simulations have correctly predicted the performances of the AGPMs (charge 2) that we have tested in the lab and used at the telescopes.[15,17]

For each numerical simulation, one needs to provide the geometrical parameters of the grating profile, which are the period, the depth, the filling factor, and the sidewall angle. In the case of the SGVC4, and with our new design using curved lines, the only parameter that is not constant is the period. The minimal period is fixed at $\Lambda^*=0.8\times\Lambda$. For the moment, the manufacturing process forbids us to further reduce the period, while keeping the depth constant. In fact, the grating trapezoidal profile still has a sidewall angle of about 3°. Shrinking the period by more than 20% would result in the merging of the grating walls, hence a reduction of the depth. The results of our simulations, performed both in the K band (2.0–2.4 μm) and in the L band (3.5–4.1 μm), are shown in Fig. 7. We computed multi-parametric maps of the theoretical mean null depth (on a logarithmic scale) as a function of the period $\Lambda$ and the filling factor $F$. The figure shows that, even if the period slightly decreases from $\Lambda$ to $0.8\times\Lambda$, and for a fixed value of the filling factor ($F=0.47$), the mean null depth remains high (blue central region) over the whole band $N\sim10^{-3}$, corresponding to a rejection ratio $R=1/N\sim1000$.

## 5. CONCLUSIONS

The SGVC2 (aka AGPM) has recently proved to be very efficient, and the manufacturing process seems mature enough to move to the next level. Increasing the topological charge of SGVCs is essential to reduce the sensitivity to low order aberrations, and will be mandatory on the future E-ELT for which a K-band instrument will not be background-limited at 1.6 $\lambda/D$ for stars brighter than $K=7$. In this paper, we have presented an original design of SGVC4 by locally varying the grating period. This new conception addresses directly the principal limitations of previous works in that domain. 2D simulations were performed with the RCWA showing a mean null depth $N\sim10^{-3}$, only slightly affected by a 20% reduction of the period. Further analyses are nevertheless necessary, since the RCWA seems not really appropriate to assess the performance of a non-symmetric SGVC, as it was for the AGPM. We are currently working on 3D rigorous numerical simulations using a finite-difference time-domain (FDTD) method, which show encouraging first results.

ACKNOWLEDGEMENT

The research leading to these results has received funding from the European Research Council under the European Union's Seventh Framework Programme (ERC Grant Agreement n.337569) and from the French Community of Belgium through an ARC grant for Concerted Research Actions.